\documentclass[modern]{aastex631}

\shorttitle{}
\shortauthors{Devillepoix et al.}

\submitjournal{MAPS}
\setwatermarkfontsize{1in}
\graphicspath{{./}{figs/}}

\begin{document}
\newcommand{\contrib}[1]{\textcolor{teal}{#1}}

\newcommand{\numb}[1]{\textcolor{orange}{#1}}
\newcommand{\source}[1]{\textsuperscript{\textcolor{blue}{[citation needed]}}\xspace}
\newcommand{\missnumber}{\numb{[NUMBER]}\xspace}
\newcommand{\checknumber}{\numb{[check number]}\xspace}

\newcommand{\fireballTimeZero}{2020-12-05THH:MM:SS.SSS}
\newcommand{\codename}{DN160822\_03}
\newcommand{\minimoon}{2019AJ....158..183S}

\title{Minimoon still on the loose}

\newcommand{\curtin}{School of Earth and Planetary Sciences, Curtin University, Perth WA 6845, Australia}
\newcommand{\adelaide}{University of Adelaide, Adelaide,  SA 5005, Australia}

\correspondingauthor{Hadrien A. R. Devillepoix}
\email{hadrien.devillepoix@curtin.edu.au}

\author[0000-0001-9226-1870]{Hadrien A. R. Devillepoix}
\affiliation{\curtin}

\author[0000-0002-8914-3264]{Seamus Anderson}
\affiliation{\curtin}

\author[0000-0002-8240-4150]{Martin C. Towner}
\affiliation{\curtin}

\author[0000-0003-4766-2098]{Patrick M. Shober}
\affiliation{\curtin}

\author[0000-0002-4079-4947]{Anthony J. T. Jull}
\affiliation{Department of Geosciences, University of Arizona, Tucson, AZ, USA}

\author[0000-0001-5390-4343]{Matthias Laubenstein}
\affiliation{INFN-Laboratori Nazionali del Gran Sasso, 67100, Assergi, AQ, Italy}

\author[0000-0003-2702-673X]{Eleanor K. Sansom}
\affiliation{\curtin}

\author[0000-0002-4681-7898]{Philip A. Bland}
\affiliation{\curtin}

\author[0000-0003-2193-0867]{Martin Cup\'ak}
\affiliation{\curtin}

\author[0000-0002-5864-105X]{Robert M. Howie}
\affiliation{\curtin}

\author[0000-0002-8646-0635]{Benjamin A. D. Hartig}
\affiliation{\curtin}

\author[0000-0000-0000-0000]{Garry N. Newsam}
\affiliation{\adelaide}

\begin{abstract}

On Aug 22, 2016, a bright fireball was observed by the Desert Fireball Network in South Australia.
Its pre-atmosphere orbit suggests it was temporarily captured by the Earth-Moon system before impact.
A search was conducted two years after the fall, and a meteorite was found after 6 days of searching. The meteorite appeared relatively fresh, had a mass consistent with fireball observation predictions, and was at the predicted location within uncertainties.
However, the meteorite did show some weathering and lacked short-lived radionuclides ($^{58}$Co, $^{54}$Mn).
A terrestrial age based on cosmogenic $^{14}$C dating was determined; the meteorite has been on the Earth's surface for $3.2\pm1.3$\,kyr, ruling out it being connected to the 2016 fireball.
Using an upper limit on the pleistieocene terrain age and the total searched area, we find that the contamination probability from another fall is $<2\%$.
Thus, the retrieval of the ``wrong" meteorite is at odds with the contamination statistics.
This is a key example to show that fireball-meteorite pairings should be carefully verified.

\end{abstract}

\section{Introduction} \label{sec:intro}

When collecting meteorites in hot and cold deserts, pairing samples together is often an issue.
This has to be carefully considered when collecting data for meteorite influx studies in particular \citep{1996MNRAS.283..551B}.
When a meteorite is found after a recently observed fall, making the connection is often simpler.
As observing a very bright fireball and finding a meteorite are both relatively low-probability events, concluding the dependence of these events is often reasonable, especially if the meteorite appears fresh (limited oxidation, dark fusion-crust, etc.).

In some cases, when a meteorite is recovered long after its fall, special attention is paid to the validity of this fireball-meteorite connection.
For instance, \citet{2014A&A...570A..39S} make a case for the connection of meteorites found in 2011 to the Bene{\v{s}}ov bolide 20 years earlier.

In this short communication, we report the case of a dedicated search of a two year old meteorite fall observed by the Desert Fireball Network in Australia. A compatible meteorite was recovered matching search criteria (predicted mass and location).
After further laboratory analyses, it was later found to have an incompatible terrestrial age.
We then discuss potential oversights when trying to pair recovered meteorites with observed fireball events

\section{Fall position prediction \& Meteorite search}

\begin{figure}
    \centering
    \includegraphics[width=1.\textwidth]{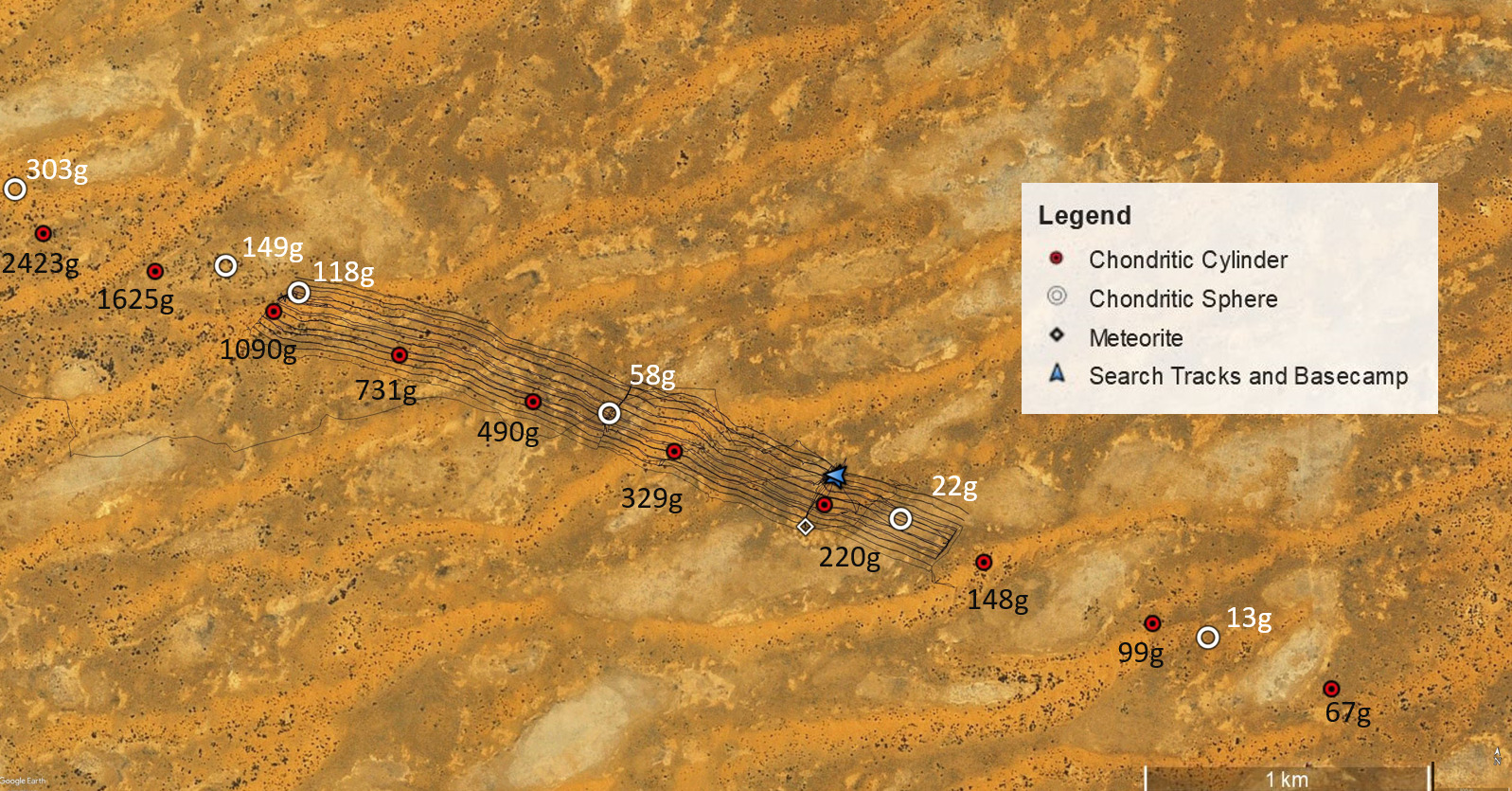}
    \caption{Predicted fall lines for chondritic sphere and cylinder, with meteorite found, overlaid with the GPS searching tracks. The location the meteorite was found is $\sim$100\,m from the nominal fall lines, and its size is consistent consistent with an object that has a slightly higher drag profile than a sphere.}
    \label{fig:fall_line}
\end{figure}

On 2016-08-22T12:17:10Z a bright fireball was observed by the Desert Fireball Network in South Australia.
It lasted 5.3\,s and was visible from 74.1\,km down to 24.1\,km.
This event stood out because of its orbit: the meteoroid had a high chance to have been temporarily captured by the Earth before impact \citep{2019AJ....158..183S}.
This trait is highly unusual, and of high interest because temporarily captured asteroids are the easiest to reach by a space probe (lowest delta-V).
The dynamical mass predictions \citep{2019ApJ...885..115S} estimate an approximately 50-100\,g surviving mass.
The dark flight trajectory of the meteorite was significantly influenced by the atmospheric winds, particularly by the subtropical jet stream (up to 90\,m\,s$^{-1}$).
Using the method of \citet{2022PSJ.....3...44T} fall lines were generated as a guide to searching for mass range of 0.04-500\,g (Fig. \ref{fig:fall_line}).
We assumed chondritic density, with a shape ranging from sphere to cylinder -- found to best match previous DFN recoveries \citep{2018M&PS...53.2212D,2020M&PS...55.2157S,2022M&PS...57.1146S,2022arXiv220206641D,2022ApJ...930L..25A}.

\begin{figure}
    \centering
    \includegraphics[width=.6\textwidth]{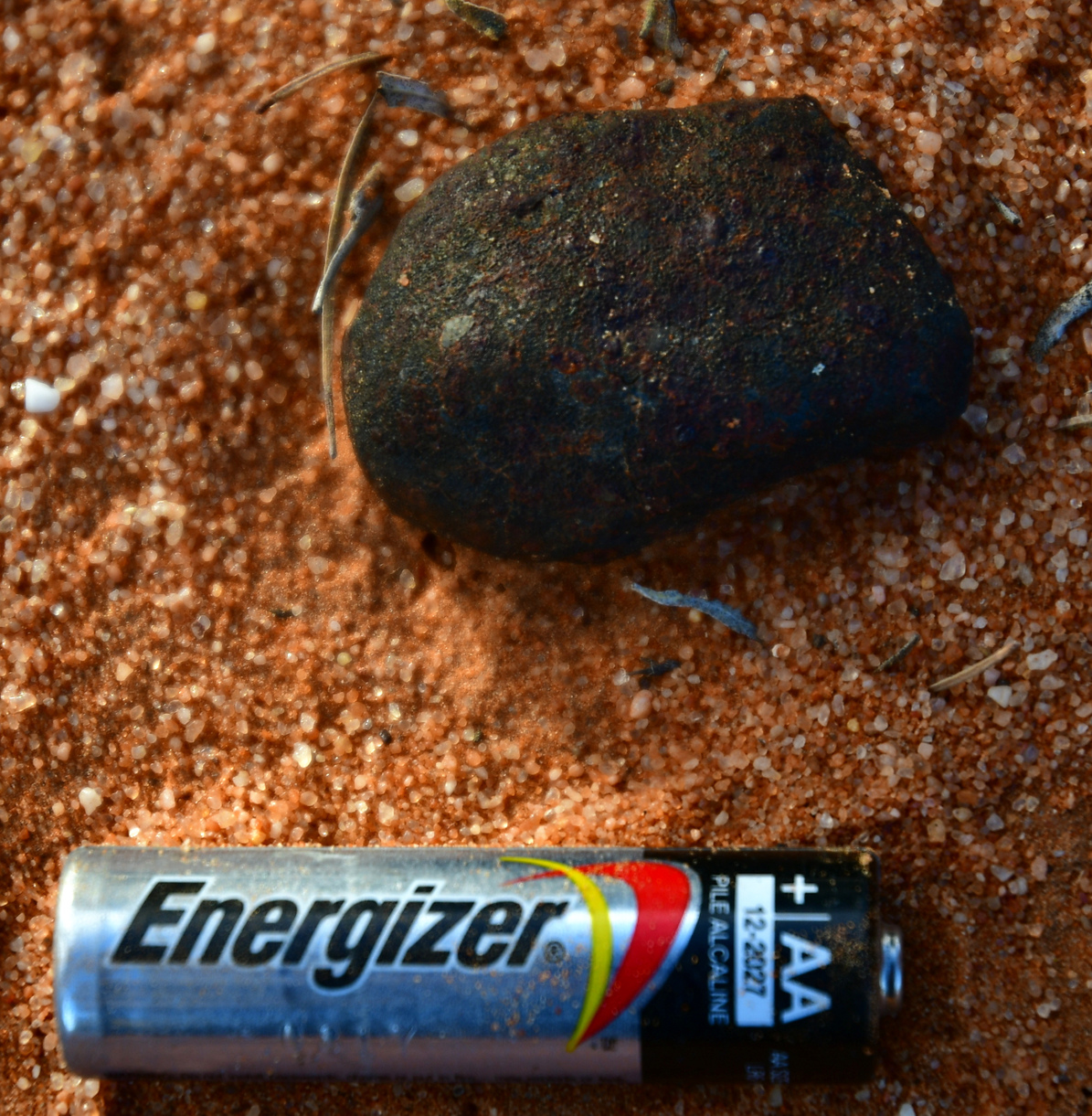}
    \caption{The Lake Frome 006 meteorite as it was found, with AA battery for scale.}
    \label{fig:meteorite}
\end{figure}

Due to the difficulty in accessing the very remote site and the relatively small size of the meteorite, this fall was not prioritised.
A 4 person search campaign was eventually launched during winter 2018, two years after the fall.
A 33.7\,g meteorite was recovered on June 3rd, 2018, after 6 days of searching, at $\phi=-30.56258$ $\lambda=140.42772$ (Fig. \ref{fig:meteorite}).
It was found $\sim$100\,m from the predicted fall line and at a point along the line corresponding to the appropriate predicted mass for a sphere (Fig. \ref{fig:fall_line}).
The meteorite's fusion crust was intact, with signs of light weathering, indicating a relatively recent fall visually consistent with the expected timeline of 2 years on the ground.
The meteorite, later classified as an L6 ordinary chondrite, was named Lake Frome 006 (MetBull 106 \footnote{\url{https://www.lpi.usra.edu/meteor/metbull.php?code=77496}}).
The team continued searching for another \numb{2} days afterwards, covering a total of \numb{0.7} km$^2$, approximately \numb{65}\% of the search area. The terrain was sparsely vegetated with very few background rocks, which were light coloured limestone.

\section{Estimate of fall statistics}

We assume the surface has been essentially stable since end of last ice age, $\sim$11500 B.C.E. \citep{Fitzsimmons}.
Given that age, and using the fall flux density of \citet{2006M&PS...41..607B}, we would expect \numb{24846} meteorite fall events of size $>\numb{30}$\,g to happen on Earth every year.
The total area searched on the campaign was $\sim$0.7\,km$^2$.
Given these constraints, the chance of another $>\numb{30}$\,g random fall on the area searched over the last 13,500 years is about 2\%.
We note that as this does not consider the weathering and potential burial of old falls, therefore this low prior probability in itself should be considered an upper bound.

\section{Short-lived radionuclides}

Freshly fallen meteorites have not experienced Earth's atmospheric protection for a significant time, and are expected to show signatures of cosmogenic radionuclides.
Measuring their concentrations can be used to assess how long samples have been on the surface -- their terrestrial age.
Cosmogenic radionuclide concentrations were analysed by means of non-destructive high purity germanium (HPGe) gamma spectroscopy.
The counting efficiencies have been calculated using thoroughly tested Monte Carlo codes.
A 32.1 g sample of Lake Frome 006 (L6 ordinary chondrite)  was measured in the underground laboratories at the Laboratori Nazionali del Gran Sasso (LNGS) \citep{ARPESELLA1996991,2017IJMPA..3243002L} for 19.09 days.
The measured $^{26}$Al activity is consistent with that expected for a small-size L chondrite \citep{2019M&PS...54..699J,2011M&PS...46..793B,2009M&PS...44.1061L}.
When we compare the radionuclide concentrations with cosmic-ray production estimations for $^{26}$Al \citep{2009M&PS...44.1061L}, and assume the specimen is from the central part, the best agreement is obtained for radii of 10-20 cm.
Otherwise it can be from the surface ($<$ 10 cm) of a bigger meteoroid.
The naturally occurring U and K (Table \ref{tab:tab2}) are well in agreement with the average concentrations in ordinary L chondrites \citep{1988RSPTA.325..535W}.
$^{228}$Th is higher than the average concentration of 43 ng g$^{-1}$.
And we do see $^{7}$Be at a level of $20 \pm 3$ mBq.
However, since no other short-lived radionuclides are visible in the spectrum, we deduce that the $^{7}$Be activity might come from some surface contamination due to fine-grained dust or soil, which is well known to contain $^{7}$Be in notable quantities.
This could also explain, why the $^{228}$Th is higher than expected, as it could be in the attached soil/dust.

\begin{table}[!ht]
    \centering
    \begin{tabular}{|l|l|l|}
    \hline
        Nuclide & Half-life (y) & Massic activity (dpm kg$^{-1})$ for \\ 
        && Lake Frome 006 (32.1 g)\\\hline
        $^{22}$Na & 2.60 & $<$ 1.4 \\ \hline
        $^{60}$Co & 5.27 & $<$ 0.79 \\ \hline
        $^{44}$Ti & 60 & $<$ 5.9 \\ \hline
        $^{26}$Al & $ 53.2 \times 10^5$ & $53.2 \pm  4.5$ \\ \hline
    \end{tabular}
    \caption{Massic activities (corrected to start of the measurement July 5th, 2018) of cosmogenic radionuclides (in dpm kg$^{-1}$) in the specimen of the Lake Frome 006 stone measured by non-destructive gamma-ray spectroscopy.
    Errors include a 1$\sigma$ uncertainty of 10\% in the detector efficiency calibration.}
    \label{tab:tab1}
\end{table}

\begin{table}[!ht]
    \centering
    \begin{tabular}{|l|l|}
    \hline
        Nuclide & Concentration in \\
        & Lake Frome 006 (21.8 g) \\ \hline
        $^{238}$U & $22.8  \pm 1.6$ ng g$^{-1}$\\ \hline
        $^{228}$Th & $76.3  \pm 4.8$ ng g$^{-1}$\\ \hline
        $^{40}$K & $700 \pm 70$ $\mu$g kg$^{-1}$\\ \hline
    \end{tabular}
    \caption{Concentration of primordial radionuclides in the specimen of the Lake Frome 006 stone measured by non-destructive gamma-ray spectroscopy. Errors include a 1$\sigma$ uncertainty of 10\% in the detector efficiency calibration.}
    \label{tab:tab2}
\end{table}

Some $^{7}$Be was detected, but as $^{58}$Co, $^{54}$Mn and $^{22}$Na were absent, the $^{7}$Be presence was most likely due to some other origin --- likely from surface contamination with fine dust or similar activated by cosmic rays in the air.
If we assume that the ratio of $^{22}$Na/$^{26}$Al (Tab. \ref{tab:tab1}) normally ranges between 1.5 and 2 for chondrites (depending on the solar activity) \citep{1994Metic..29..443B}, then 
one can exclude that the fall is younger than 13-16 years.

\section{Terrestrial age}

The terrestrial age was estimated from $^{14}$C extracted from the meteorite using the procedure of \citet{1993Metic..28..188J,2010M&PS...45.1271J}.
A sample of $\sim$0.25 g of the meteorite was crushed and cleaned with 85\% H$_{3}$PO$_{4}$ for one hour.
The cleaned sample was washed with distilled water and dried in air in an oven.
The cleaned powder is mixed with $\sim$5 g of iron chips used as a combustion accelerator and placed in a ceramic crucible.
The crucible and sample are preheated to 500 \degr C in air to remove organic contaminants.
The cleaned sample (0.231 g) and crucible are then placed in an RF furnace and the sample is rapidly heated to melting in 2 minutes by RF induction in a flow of oxygen.
The evolved gas is collected at -196\degr C and the CO$_{2}$ is separated from oxygen and other incondensable gases by vacuum distillation.
The CO$_{2}$ is converted to graphite \citep{slota1987preparation} and the graphite powder is pressed into an accelerator target and $^{14}$C is determined by accelerator mass spectrometry using the 2.5MV Pelletron AMS at the University of Arizona.
The $^{14}$C content was determined to be $34.7\pm 0.2$ dpm/kg and from this, we can calculate a terrestrial age of $3.2\pm  1.3$ kyr for an L6 meteorite with a pre-atmospheric radius of $>$20 cm.
If the sample can be constrained as $<$15cm radius, this could be consistent with a more recently fallen object \citep{1996M&PS...31..265W} in the last 1 kyr, but due to the absence of short-lived nuclides, it cannot be a ``fresh" fall.

\section{Discussion}

Despite the appearance of a fresh fall, in the correct area, Lake Frome 006 is not the meteorite recently observed by the DFN. 
Had the presence of short-lived radio-nuclides or the terrestrial age not been measured, it could have been tempting to put the mild degree of weathering down to the two years it spent on the ground.
With a $<2\%$ chance of another fall contaminating the search site, we could also have erroneously concluded that the recovered meteorite is likely connected to the fireball observed in 2016 at a $>2\sigma$ confidence.
Without other geo-chemical or cirmunstancial evidence, the fireball-meteorite connection hinges on a low contamination probability.
Establishing this prior probability relies on objective constraints put on meteorite accumulation on the area searched.
Here we use a geological argument to put an upper limit on the age of the surface, but  determining how long a meteorite could have survived on the surface in general is not trivial.
The definition of search area can also be a source of confusion.
Here we used the total area combed during the 8 day campaign, not merely the area searched until the meteorite discovery on the 6th day.
Although this would have lowered the contamination probability further, not including the full search area is flawed from a Bayesian point of view, because we would be using posterior information to inform our prior contamination probability.
Even with a convincingly low contamination probability, getting cosmogenic nuclide evidence of low terrestrial age is a worthwhile endeavour.

Amongst the $\sim$50 meteorites recovered from instrumentally observed fireballs, not all have have indisputable circumstantial evidence of "freshness" or been subject to routine geochemical validations.
The only meteorite fragments to have been recovered more than a year after their suspected fall, were those recovered in 2011 and claimed to be connected to the Bene{\v{s}}ov bolide in 1991 \citep{2014A&A...570A..39S}.
To support their claim, \citet{2014A&A...570A..39S} argue the meteorites could not have been on Earth for more than a few decades because of the degree of weathering. From this they make the argument that it is statistically impossible ($10^{-5}$) that another fall happened recently on the fall area ($250 \times 70$ m $=0.0175$\,km$^2$).
In this instance, the assumption of how long a meteorite could have taken to disappear within the area is not quantified or backed by a weathering study.
The probabilistic argument of \citet{2014A&A...570A..39S} is also highly dependent on the definition of the search area.
The authors a posteriori defined the search area as a rectangular area bounding the location of samples found, and not the total area searched for this fall.
Yet one cannot define a prior probability with posterior information \citep{1763RSPT...53..370B}.
A more appropriate choice for the normalisation area should be the total area combed, which was not reported in the study.
Finally the authors support their claim by saying that the spectrum of the fireball was consistent with a chondritic composition (the same as the meteorites).
75\% of meteorite falls being chondritic, the spectrum-composition match is hardly indicative in this case.
Although Bene{\v{s}}ov is the only case recovered more than a year after its fall, it may not be the only case where a fireball-meteorite pairing could come into question.
The instrumentally observed meteorite fall with the next longest recovery time was Mason Gully \citep{2016M&PS...51..596D}.
It was recovered $>6$ months after its fall, it also appears slightly weathered, and no short-lived radio-nuclides or terrestrial dating have not been reported.

\section{Conclusions}

The issue of fireball-meteorite parings can in part be treated statistically --- by answering the question: "What are the chances that the meteorites found are from another fall?" --- but it is riddled with subjectivity around the prior assumptions used.
We recovered a meteorite within the predicted mass range and fall area of an instrumentally observed fall from 2016.
Despite the low contamination probability from another older fall ($<2\%$), we have performed terrestrial dating and found it to be an old fall ($3.2\pm  1.3$ kyr).
About \numb{35}\% of the fall area for the 2016 fireball remains to be searched, therefore the meteorite could still plausibly be uncovered by a future search party.
This is a key example to show that fireball-meteorite pairings should be carefully verified.
As the number of meteorites with precisely measured orbits increase, the likelihood of meteorite-fireball pairings should be rigorously considered.
Terrestrial dating and short-lived radionuclide testing are geo-chemical techniques that should be more generally adopted to confirm fresh falls.

\begin{acknowledgements}

This work was funded by the Australian Research Council as part of the Australian Discovery Project scheme (DP170102529, DP200102073), the Linkage Infrastructure, Equipment and Facilities scheme (LE170100106), receives institutional support from Curtin University. John Whitehead is thanked for his work during the searching expedition.

\end{acknowledgements}

\bibliography{biblio}
\bibliographystyle{aasjournal}

\end{document}